\documentclass[conference,letterpaper]{IEEEtran}

\usepackage[utf8]{inputenc} 
\usepackage[T1]{fontenc}
\usepackage{url}
\usepackage{ifthen}
\usepackage{cite}
\usepackage[cmex10]{amsmath} 

\usepackage{pgfplots}
\usepgfplotslibrary{groupplots,dateplot}
\usetikzlibrary{patterns,shapes.arrows}
\pgfplotsset{compat=newest}
\usepackage{tikzscale}
\usepackage{balance}
\usepackage{tabularx} 
\usepackage{amsmath, amssymb, amsthm}  
\usepackage{graphicx} 
\usepackage{cite} 
\usepackage[final]{hyperref} 
\usepackage{bbm}
\hypersetup{
	colorlinks=true,       
	linkcolor=blue,        
	citecolor=blue,        
	filecolor=magenta,     
	urlcolor=blue         
}
\usepackage{blindtext}
\usepackage{enumitem}
\usepackage{subcaption}
\newcommand\numberthis{\addtocounter{equation}{1}\tag{\theequation}}
\newcommand{\U}{\mathbf{U}}
\newcommand{\X}{\mathbf{X}}
\newcommand{\x}{\mathbf{x}}
\newcommand{\Y}{\mathbf{Y}}
\newcommand{\y}{\mathbf{y}}
\newcommand{\M}{\mathbf{M}}
\newcommand{\m}{\mathbf{m}}
\newcommand{\Z}{\mathbf{Z}}
\newcommand{\z}{\mathbf{z}}

\renewcommand{\S}{\mathbf{S}}
\newcommand{\s}{\mathbf{s}}
\newcommand{\pins}{p_{\mathrm{ins}}}
\newcommand{\pdel}{p_{\mathrm{del}}}
\newcommand{\psub}{p_{\mathrm{sub}}}
\newcommand{\prep}{p_{\mathrm{cor}}}

\newcommand{\bb}{{\beta_{\text{b}}}}
\newcommand{\bi}{{\beta_{\text{i}}}}
\newcommand{\be}{{\beta_{\text{e}}}}
\newcommand{\bo}{{\beta_{\text{o}}}}

\DeclareMathOperator*{\argmax}{arg\,max}
\DeclareMathOperator*{\head}{f^{from}}
\DeclareMathOperator*{\tail}{f^{to}}
\DeclareMathOperator*{\elabel}{f^{lbl}}

\newif\ifnotes
\notesfalse
\newcommand{\gnote}[1]{\ifnotes{{\sf\color{blue} [Gopi: #1]}}\fi}
\newcommand{\hnote}[1]{\ifnotes{{\sf\color{red} [Henry: #1]}}\fi}

\begin{document}
\title{Trellis BMA: Coded Trace Reconstruction \\ on IDS Channels for DNA Storage} 


\author{%
  \IEEEauthorblockN{Sundara Rajan Srinivasavaradhan\IEEEauthorrefmark{1},
                    Sivakanth Gopi\IEEEauthorrefmark{2},
                    Henry D. Pfister\IEEEauthorrefmark{3}\IEEEauthorrefmark{2},
                    and Sergey Yekhanin\IEEEauthorrefmark{2}}
  \IEEEauthorblockA{\IEEEauthorrefmark{1}%
                    University of California, Los Angeles, Electrical and Computer Engineering, sundar@ucla.edu}
  \IEEEauthorblockA{\IEEEauthorrefmark{2}%
                    Microsoft Research, Redmond, WA, \{sigopi,yekhanin\}@microsoft.com}
  \IEEEauthorblockA{\IEEEauthorrefmark{3}%
                    Duke University, Electrical and Computer Engineering, Durham, NC, henry.pfister@duke.edu}
}

\maketitle

\thispagestyle{plain}
\pagestyle{plain}

\begin{abstract}
Sequencing a DNA strand, as part of the read process in DNA storage, produces multiple noisy copies which can be combined to produce better estimates of the original strand; this is called \emph{trace reconstruction}. One can reduce the error rate further by introducing redundancy in the write sequence and this is called \emph{coded trace reconstruction}. 
In this paper, we model the DNA storage channel as an insertion-deletion-substitution (IDS) channel and design both encoding schemes and \emph{low-complexity} decoding algorithms for coded trace reconstruction.

We introduce Trellis BMA, a new reconstruction algorithm whose complexity is linear in the number of traces, and compare its performance to previous algorithms.
Our results show that it reduces the error rate on both simulated and experimental data.
The performance comparisons in this paper are based on a new dataset of traces that will be publicly released with the paper.
Our hope is that this dataset will enable research progress by allowing objective comparisons between candidate algorithms.


\end{abstract}


\section{Introduction}

\begin{figure*}[!h]
\centering
\includegraphics[width=2\columnwidth]{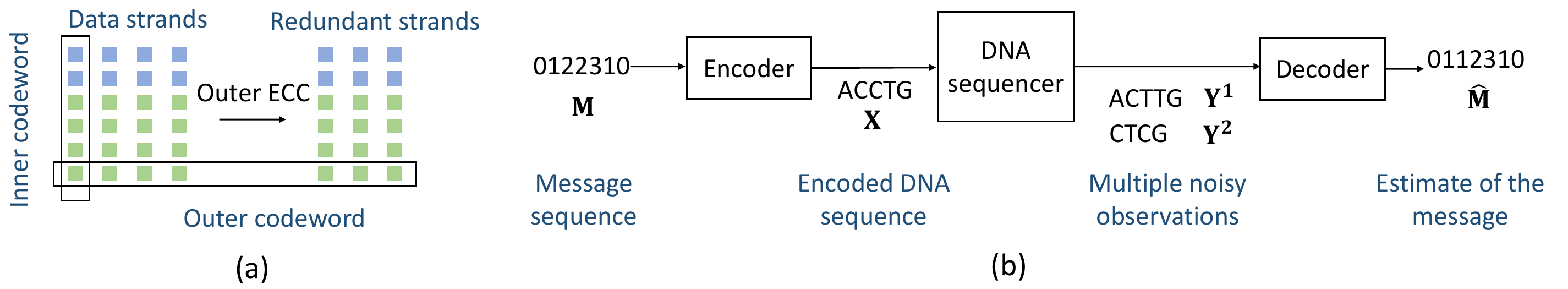}
\vspace{-1mm}
\caption{\small (a) The interplay between inner and outer code in a DNA storage system. Data strands are first encoded using an outer code (to correct for missing sequences) and then using an inner code which corrects IDS errors. (b) The inner code architecture for DNA storage.  Encoded DNA strands are read or ``sequenced'' using a sequencing technology, such as Illumina/Nanopore sequencers, and this outputs many noisy  copies of the DNA sequence, from which the message vector in the data strand is recovered.
 \label{fig:dna_storage}}
 \vspace{-4mm}
\end{figure*}

DNA storage is an exciting area because of its potential  to provide both high information density and long-term stability~\cite{Church-science12}. To achieve a good trade-off between efficiency and reliability, DNA storage systems use error-correcting codes~\cite{Grass-acie15,Yazdi-mbmc15,Yazdi-sr15,Heckel-isit17,Yazdi-nature17,Organick-natbiotech18,Lenz-icassp20,Antkowiak-naturecomm20}. This paper considers the design and decoding of error-correction codes for the DNA storage channel (see Figure~\ref{fig:dna_storage}). 

In this paper, the DNA storage channel is modeled as an insertion-deletion-substitution (IDS) channel and we focus on the case where a single encoded message is transmitted and multiple independent \textit{traces} are observed~\cite{Grass-acie15,Yazdi-mbmc15,Yazdi-sr15,Organick-natbiotech18}.
Sequence reconstruction methods for this problem date back to the 1980s~\cite{Carrillo-siamam88}. 
This is closely related to the trace reconstruction problem in CS literature which asks how many traces (from a deletion channel) are needed to perfectly reconstruct the input message sequence, in the average or worst case. Many algorithms exist for trace reconstruction~\cite{Levenshtein-it01,Batu-soda04,TR-SODA08,ADS-17,NP-17,HPP-2018,Chase-20}, a few of which (such as \textit{Bitwise Majority Alignment} (BMA)~\cite{Batu-soda04}) can be modified for the IDS channel and have been used in DNA data storage systems~\cite{BMA-18, BMA-20}.

In practical systems, outer codes are used to code across multiple DNA strands in order to recover missing sequences and correct substitutions of individual symbols.
Thus, we focus primarily on approximate reconstruction, as opposed to exact reconstruction.
For IDS-like channels, one can compute exact posterior marginals by combining ideas from multiple-sequence alignment~\cite{Carrillo-siamam88} and the BCJR algorithm~\cite{Bahl-it74} (e.g., see~\cite{Davey-it01,Sakogawa-isit20,Lenz-itw20}).
Using these posterior marginals, it is easy to compute estimates that minimize additive distortion measures.
If the outer code uses hard-decision decoding, then a reasonable goal is to construct a practical estimator that, given a small number of traces, minimizes the expected Hamming distance to the input message.
Strands may also use an inner code that is designed to provide additional protection~\cite{CGMR-20,BLS-20}.
The inner code constraints can also be included in channel trellis~\cite{Chandak-icassp20} so that trellis-based methods can still be used for inference.
In particular, for convolutional codes, it is possible to build a multidimensional trellis and perform symbolwise maximum-a-posteriori (MAP) reconstruction, as observed in~\cite{Lenz-itw20}. But, the complexity grows exponentially with the number of traces making exact inference infeasible.


\vspace{-0mm}

\subsection{Contributions\protect\footnote{The majority of this work was completed while the first author was an intern at Microsoft Research and was presented there on Sept.\ 11th, 2020.}}
\begin{itemize}[leftmargin = 3mm]
    \item A low-complexity heuristic dubbed Trellis BMA is proposed that allows multiple single-trace trellis decoders to interact and estimate the input message on-the-fly. This is different from the approaches in~\cite{srinivasavaradhan2019symbolwise, Srinivasavaradhan-arxiv20,Lenz-itw20} because each single-trace trellis decoder is influenced by the other decoders but it is related to the factor graph method in~\cite{Sakogawa-isit20}. Our idea marries BCJR inference~\cite{Bahl-it74} for IDS channels~\cite{Davey-it01} with the consensus approach of BMA, hence the name \emph{Trellis BMA}.
    
    \item A dataset of short strand DNA reads is generated that can be used to compare algorithms with actual DNA reads. This dataset will be released publicly to serve as a benchmark for coded trace reconstruction algorithms.
    
    
    \item A new construction for the multi-trace IDS trellis is provided where the number of edges grows at a lower exponential rate (with the number of traces) than previous approaches. Using BCJR inference to compute the symbolwise posterior probabilities for multiple traces is exponentially faster with this formulation. 

\end{itemize}


\vspace{-0mm}
\section{Background}

\subsection{DNA sequencing channel}
The observed noise in DNA storage is a complicated combination of synthesis errors, amplification errors, and sequencing noise~\cite{Mao-it18}.  
Even if we ignore the first two elements, the exact error profile of the noisy observations is dependent on the DNA sequencing technology used.
However, exactly modeling this error profile is tedious and often impractical. Moreover, DNA sequencing technologies are evolving at a rapid pace and focusing on a particular error profile does not provide a future-proof approach to the problem.
Instead, one typically considers a simplistic approximation and models the sequencing channel as an IDS channel (defined in the next subsection). Our ideas also extend naturally to more complex approximations for the channel model. For instance, insertions and deletions often occur in ``bursts'' and such events can be captured by a 
first-order Markov model; our decoder can easily be modified to accommodate for such variations.

Due to the difficulty of synthesizing and sequencing long DNA strands, DNA storage systems typically encode a single file into many different short strands.
The Poisson nature of sampling short strands from the pool means that many of these strands will not be sequenced.
Thus, an outer code is required and sequence numbers must be included for disambiguation~\cite{Shomorony-it21}.
This detail is sometimes neglected in simulation-based experiments (e.g., it seems a single long strand is used in~\cite{Lenz-itw20}).

\subsection{Insertion deletion substitution channel}
The insertion deletion substitution (IDS) channel is defined by its input/output alphabet $\Sigma$ and four non-negative parameters $\pins,\ \pdel,\ \psub,\ \prep$
with $\pins+\pdel+\psub+\prep=1$. Given an $N$-length input sequence $\X=X_1X_2...X_N \in \Sigma^N$, the IDS channel sequentially takes in one input symbol at a time
and constructs a variable length output $\Y = Y_1Y_2... \in \Sigma^*$ sequentially, where $\Sigma^* \triangleq \cup_{m=0}^{\infty} \Sigma^m$ is the set of finite strings over $\Sigma$. Let the input pointer be $i$ and the output pointer be $j$.
Starting from $i=j=1$, sample from the following events until $i$ equals $N+1$: \\
$\bullet$ \textbf{Insertion} (probability $\pins$): choose $Y_j$ uniformly at random from $\Sigma$, increase $j$ by 1,  and leave $i$ unchanged;\\
$\bullet$ \textbf{Deletion} (probability $\pdel$): increase $i$ by 1 and leave $j$ unchanged;\\
$\bullet$ \textbf{Substitution} (probability $\psub$): choose $Y_j$ uniformly at random from $\Sigma\setminus \{X_i\}$ . Increase both $i$ and $j$ by 1;\\
$\bullet$ \textbf{Correct:} (probability $\prep$): Set $Y_j=X_i$. Increase $i$ and $j$ by 1.

\begin{table}
\centering
\renewcommand{\arraystretch}{1.2}
\begin{tabular}{ | m{3.6cm} | m{4.0cm}| } 
\hline
\multicolumn{2}{|c|}{\bf Notation and acronym quick reference} \\
\hline
\hline
IDS  &  Insertion Deletion Substitution \\ 
\hline
Trace & Output of the IDS channel \\
\hline
$\Sigma$ & IDS channel input / output alphabet \\
\hline
Upper-case letters (e.g. $X$) & random variable or integer constant (should be clear based on context) \\ 
\hline
Lower-case letters (e.g. $x$) & generic variable \\ 
\hline
Bold-face letters (e.g. $\x$) & sequence or vector \\ 
\hline
Bold upper-case letters (e.g. $\X$) & random vector \\ 
\hline
Subscripts (e.g. $x_n$) & $n$-th symbol of sequence $\x$ \\ 
\hline
Superscripts (e.g. $\Y^k$) & $k$-th trace \\ 
\hline
Superscript range (e.g. $\Y^{1:K}$) & Tuple of traces $(\Y^1,\Y^2,...,\Y^K)$ \\ 
\hline
TR & Trace reconstruction \\ 
\hline
MAP & Maximum a-posteriori \\ 
\hline
BMA & bitwise majority alignment \\ 
\hline
Trellis BMA & Trellis bitwise majority alignment\\
\hline
Improved BMALA & Improved BMA with lookahead \\ 
\hline
CC & Convolutional code \\ 
\hline
FSM & Finite-state machine\\
\hline
MR Code & Marker repeat code\\
\hline
\end{tabular}
\end{table}

\vspace{-1mm}
\subsection{Trace reconstruction with and without coding}
As discussed in the introduction, the trace reconstruction (TR) problem has been formalized in the CS literature as the question, ``How many traces are required to exactly reconstruct $\X$?''~\cite{Batu-soda04,TR-SODA08,ADS-17,NP-17,HPP-2018,Chase-20}. However, exact reconstruction is typically impossible from only a few traces~\cite{Chase-ihp21}.
Thus, we use the term \emph{TR algorithm} for any algorithm that uses multiple independent traces $\Y^1,\Y^2,...,\Y^K \in \Sigma^*$ of $\X$ to construct an estimate $\widehat{\X}(\Y^{1:K})$, where $\Y^{1:K}$ is shorthand for $(\Y^1,\Y^2,...,\Y^K)$.

A more general formulation is to consider a code that maps a message sequence $\M = M_1M_2...M_L \in \mathcal{M}^L$ to a  codeword $\X = X_1X_2...X_N \in \Sigma^N$. The goal of coded TR is to compute an estimate $\widehat{\M}(\Y^{1:K})$  
of the message sequence $\M$ from tke $K$ multiple independent traces $\Y^1,\Y^2,...,\Y^K$ of $\X$. This setup 
naturally fits the DNA storage architecture in Fig.~\ref{fig:dna_storage}.

\subsection{Error-correcting codes}

For the inner code, this work considers marker repeat (MR) codes with the addition of a random scrambling vector to prevent shift invariance.
Marker codes are synchronization codes where a short \emph{marker sequence} is inserted periodically~\cite{Ratzer-anntele05}.
MR codes are a new variation where, periodically, a single input symbol is transmitted multiple times.
For example, a length-$N$ MR code with $r$ length-2 repeats satisfies $x_{n+1}=x_{n}$ when $n=\lfloor iN/(r+1) \rfloor$ for $i=1,\ldots,r$.
Results are given for MR codes with $N=110$ and $r=6,10$.
Rate-1/2 quaternary convolutional codes with memory 3-5 and puncturing were also tested and found to be inferior to MR codes above rate 3/4 (see Appendix~\ref{app:MR-vs-CC}).


While this work focuses on the efficient decoding of the inner code when multiple traces are received, our analysis also assumes there will be an outer code.
In particular, we target schemes where the inner codes are decoded first followed by the outer code.
In contrast to~\cite{Lenz-itw20}, we do not consider iteration between the inner and outer decoder nor do we estimate the error rate after decoding of the outer code.

\subsection{Performance metrics and information rates}

The choice of performance metric for BCJR inference depends crucially on how the outputs will be used.
Different decoding methods for the outer code lead to different achievable rates.
Any rate loss due to inner MR codes is included in these computations whereas rate loss due to sequence numbers, which are typically required by outer codes, is neglected.

For general trace reconstruction (or detection before hard-input decoding of an outer code defined over $\Sigma$), one typically chooses $\widehat{\X}$ to minimize the expected Hamming distance \vspace{-1mm}
\begin{equation}
\mathbb{E}\left[ d(\widehat{X},X) \right] = \sum_{n=1}^N \Pr(\widehat{X}_n \neq X_n),
\label{eq:TR_smap} \vspace*{-1mm}
\end{equation}
and the optimal $\widehat{X}$ is given by the symbolwise MAP estimate.
Choosing $\widehat{\X}$ to minimize the edit distance has also been considered in~\cite{Bio-2020,DRRS-20,Srinivasavaradhan-arxiv20}.
For hard-decision decoding of an outer code defined by $M$ symbols, the expected Hamming error rate is likewise minimized by choosing $\widehat{\M}$ to be the symbolwise MAP estimate of $\M$.

For soft-decision decoding, the outer decoder uses the posterior marginals, $U_{l}(m) \triangleq \Pr(M_l \!= \! m|\Y^{1:K})$, whose uncertainty is quantified by the average symbolwise entropy
\begin{equation}
H = \frac{1}{L} \sum_{l=1}^L \mathbb{E} \left[ \frac{1}{\log U_l (M_l)} \right] \leq \frac{1}{L} \sum_{l=1}^L \mathbb{E} \left[ \log \frac{1}{\widehat{U}_l (M_l)} \right].
\label{eq:CodedTR_smap}
\end{equation}
Here, $\widehat{U}_l$ is any approximate posterior marginal (e.g., due to channel mismatch or suboptimal processing) satisfying $\sum_{m\in \Sigma} \hat{U}_l
(m) \!=\! 1$ for all $l$.
For i.i.d. equiprobable inputs into a rate-$R$ inner code, the quantity $(2-H)R$ (bits/base) is an overall achievable information rate (AIR) for separate detection and decoding, called the BCJR-once rate~\cite{Kavcic-it03,Muller-it04,Soriaga-it07}.
If a random outer code is used with joint decoding, then the AIR is the mutual information rate $\frac{1}{N}I(\M;\Y^{1:K})\!=\!\frac{R}{L}I(\M;\Y^{1:K})$ which can be estimated using the BCJR algorithm~\cite{Pfister-globe01,Arnold-it06,Kavcic-isit04}.

In actual DNA storage systems, the number of traces $K$ will be a random variable that is different for each observed cluster.
In that case, a particular AIR for random $K$ is given by averaging that AIR over the distribution of $K$.

\section{Dataset} \label{sec:dataset}

The performance comparisons in this paper are based on a new dataset of 269,709 traces of 10,000 uniform random DNA sequences of length $110$ that is now publicly available at: \vspace{-1mm}
\begin{center}$\!\!\!$\url{https://github.com/microsoft/clustered-nanopore-reads-dataset}$\!\!\!\!$\end{center} \vspace{-1mm}
Our hope is that this dataset will enable further research progress by allowing objective comparisons between the algorithms. DNA sequences were synthesized by Twist Bioscience and amplified using polymerase chain reaction. The amplified products were ligated to Oxford Nanopore Technologies (ONT) sequencing adapters by following the manufacturer's protocol (LQK-LSK $109$ kit). Finally, ligated samples were sequenced using ONT MinION. Clusters of noisy reads have been recovered using the algorithm from~\cite{NIPS-17}.
The insertion, deletion, and substitution rates for this dataset are roughly $\pins = 0.017$, $\pdel = 0.02$, and $\psub = 0.022$.

\vspace{0.75mm}
\noindent
\textbf{Note added on 8/12/2024:}
We would like to thank Adar Hadad who pointed out to us that the collection of 10,000 DNA sequences of length 110 generated for this study exhibits long-range dependencies instead of being uniformly random. This is due to an error in the generation process. Since the input sequences are not uniform, the clustering algorithm from~\cite{NIPS-17} may have unexpected behavior and some recovered clusters may be malformed, making the trace reconstruction problem harder.

\vspace{0.75mm}
\noindent
\textbf{Using the dataset for coded TR:} The dataset is a collection of $(\x,\y)$ pairs allowing one to estimate the expected performance of TR algorithms for uniform random DNA sequences. For coded TR, the problem is that one cannot estimate an expectation over codewords because the randomly generated DNA sequences are unlikely to be codewords in the code.

One can estimate the expected performance for a coded system with random scrambling.
Assume $\Sigma$ has an abelian group structure and
let the code $\mathcal{C}\subseteq \Sigma^N$ be a subset with encoder $\mathcal{E}\colon \mathcal{M}^L \to \mathcal{C}$. 
Consider estimating a performance measure $\phi = \mathbb{E}[\Phi (\Y^{1:K};\M,\Z)]$ for the scrambled encoder defined by $\X = \mathcal{E}(\M)+\Z$, where $\Z \in \Sigma^N$ is a uniform random scrambling sequence.
Since this induces a uniform distribution on $\X$ (see Appendix \ref{app:random_scrabling} for a proof), the dataset can be used to estimate $\phi$.
For an $\x$ in the dataset, let $T(\x)$ denote the set of $\y$ traces generated by $\x$. 
Samples can be drawn as follows:

\begin{itemize}[leftmargin = 3mm]
    \item Let $\x$ be the result of drawing a uniform random DNA sequence from the dataset, $\m$ be the result of choosing a uniform random message, and then compute $\z = \x - \mathcal{E}(\m)$.
    \item Compute the sample value $\Phi(\y^{1:K};\m,\z)$ for $K$ traces sampled randomly from $T(\x)$ without replacement.
\end{itemize}
\gnote{Consider adding this sentence if we have space: "To summarize, we can find the expectation of the metric $\Phi$ 
over the ensemble of codes given by $\{\mathcal{C}+z: z\in \Sigma^N\}$, and therefore we expect that the metric for a random code from this ensemble is close to the expected value with high probability."}

\noindent
To summarize, for an encoder $\mathcal{E}$, we estimate the average of $\Phi$ over $\Z$.
Hence, there is a $\z$ that performs this well or better.
In some cases, one might also expect the value of $\Phi$ to concentrate around its expectation and establishing this (e.g., sufficient conditions) is an interesting open question.



\hnote{Maybe add proof for this in appendix?}

\section{Algorithms for TR and Coded TR}
\subsection{Multi-trace trellis via hidden Markov model}
\label{subset:multiD}
Our discussion of algorithms begins with a brief description of a hidden Markov model (HMM) associated with the problem.
The state diagram of this HMM implies a natural multi-trace IDS trellis that is different from previous methods \cite{Davey-it01,Sakogawa-isit20, Lenz-itw20}.
This trellis has significantly fewer edges and this reduces the complexity of BCJR inference.
However, the resulting trellis and BCJR definitions are a bit different from those typically used in coding theory.
We refer the reader to Appendix~\ref{app:prob_fsm} for a detailed description of trellis and BCJR inference.

In essence, our construction of the trellis describing the joint distribution of $(\M, \X, \Y^1,\Y^2,...,\Y^K)$ avoids local exponential blow-up in the number of edges by
\begin{itemize}[leftmargin = 3mm]
    \item modeling insertion events as vertical edges, thereby sequentially accounting for insertions.
    \item modeling events in each trace sequentially.
\end{itemize}
Consider a message sequence $\M=M_1M_2...M_L$, where $M_l \in \mathcal M$,  which is mapped onto a codeword $\X=X_1X_2...X_N$, where $X_i \in \mathcal X$, using a (possibly time-varying) deterministic FSM encoder. 
Such an encoder takes as input a message symbol $M_i$, transitions to state $Q_i$ and emits $u$ codeword symbols $X_{u(i-1)+1}X_{u(i-1)+1}...X_{ui}$. The transition and codeword symbols emitted only depend on $M_i$ and its state $Q_{i-1}$ before accepting input symbol $M_i$.  For simplicity, assume that the number of emitted symbols $u$ is fixed for all $i$ ($N=Lu$); our trellis can also account for cases where $u$ varies with $i$.

Suppose we observe $K$ independent traces $\y^1,...,\y^K$ generated from $\X$. Let $\y^k=y^k_1y^k_2...y^k_{R_k}$, therefore the length of the $k$-th trace is $R_k$. The trellis is a directed acyclic graph (DAG) with weighted edges where the vertices are ordered by ``stages'' -- edges connect two vertices in the same stage or connect a vertex at stage $t$ to a vertex at stage $t+1$.
We note that generalizes the standard notion of a trellis by allowing edges between vertices in the same stage. 
At stage $t$, vertex $v_t$ is defined by $(q_t, p^1_t, p^2_t,...,p^K_t, m_t, x_t)$ where
\begin{itemize}[leftmargin = 2mm]
    \item $q_t \in \mathcal Q$, is the state of the encoder at stage $t$;
    \item $p^k_t \in \mathcal P^k_t$ with $\mathcal P^k_t = \{1,2,...,R_k\}$ is the output pointer which, for the $k$-th trace at stage $t$, equals the index of the output currently being explained;
    \item $m_t {\in} \mathcal M_t$ with $\mathcal M_t {=} \mathcal M \cup \{\star\}$, is the on-deck message symbol;
    \item $x_t {\in} \mathcal X_t$ with $\mathcal X_t {=} \mathcal X \cup \{\star\}$, is the on-deck codeword symbol.
\end{itemize}

Therefore, $v_t\in \mathcal \mathcal Q \times \mathcal P^1_t \times \mathcal P^2_t ... \times \mathcal P^K_t \times \mathcal M_t \times \mathcal X_t$, where $\times$ denotes the Cartesian product. For clarity, we construct the trellis stage-by-stage, describing the stages corresponding to the first message symbol.\\
\textbf{Modeling the input.} 
An edge connects vertex $v_1= (q_{init},1,1,...,1,\star,\star)$ at stage 1 to $v_2=(q,1,1,...,1,m,x)$ at stage 2, where $q_{init}$ is the initial state of the encoder and encoder makes the transition $q_{init}\rightarrow q$ when presented with input $m$, emitting first codeword symbol $x$. The edge weight is equal to $\Pr(M_i=m)$ to  model the input distribution.\\
\textbf{Modeling IDS events.} 
An edge connects a vertex $v_2=(q,p^1,p^2,...,p^K,m,x)$ to $v_3=(q,p^1,p^2,...,p^K,m,x)$ with a weight equal to $\pdel$ modeling a deletion event in the first trace. An edge connects a vertex $v_2=(q,p^1,p^2,...,p^K,m,x)$ to $v_3=(q,p^1+1,p^2,...,p^K,m,x)$ with a weight equal to $\prep$ if $y^1_{p^1}=x$ and $\frac{\psub}{|\mathcal X|-1}$ otherwise.
This models a substitution/correct event in the first trace. An edge connects a vertex $v_2=(q,p^1,p^2,...,p^K,m,x)$ to $v'_2=(q,p^1+1,p^2,...,p^K,m,x)$ in the same stage with a weight equal to $\frac{\pins}{|\mathcal X|}$, modeling an insertion event in the first trace. Notice how only the output pointer to the first trace changes in all cases. We construct $K$ such stages for $K$ traces.\\
\textbf{Updating on-deck codeword symbol.} We have only considered the events corresponding to the first codeword symbol so far.  Next, we update the output buffer to replace the first codeword symbol $x$ by the second $x'$, followed by $K$ stages of IDS event modeling for the second codeword symbol.\\
\textbf{Transitioning to the next input.} The above two steps of modeling the IDS events and updating the output buffer are repeated until all codeword symbols for a given input symbol are processed. Then, the input and output buffer are cleared and the next message symbol is accepted.

The above steps comprise one input cycle. These steps are repeated until all message symbols are exhausted. 
Each path connecting $(q_{init},1,1,...,1,\star,\star)$ at the first stage to $(q_{end},R_1,R_2,...,R_K,\star,\star)$ at the final stage correspond to a message  sequence and a sequence of events that resulted in the observed traces $\Y^1=\y^1,...,\Y^K=\y^K$. The weight of this path is the joint probability of observing the message, the sequence of events and the traces. For this setup, one can use BCJR inference to compute the posterior probability that the true system passed through a given vertex at a particular stage. Then, one can compute $\Pr(M_l=m|\Y^1=\y^1,...,\Y^K=\y^K)$ by summing the posterior probabilities of all vertices associated with message symbol $m$ in the input cycle of stage $l$.

\textbf{Time Complexity.} Assuming the length of the traces $R_k=O(N)\ \forall k$, and $\mathcal Q$ is the state-space of the encoder FSM, the total number of edges in the trellis is $O(N^{K+1}K|\mathcal Q|)$, which is the time complexity to exactly compute the APPs. In practice, it is reasonable to assume that the output pointer does not drift too far from the input pointer for each IDS channel, i.e., at a given stage one assumes that $|\mathcal P_t|=\Delta < N$~\cite{Davey-it01,Sakogawa-isit20,Srinivasavaradhan-arxiv20}.
Using this assumption, the complexity to compute APPs is roughly $O(NK|\mathcal Q|\Delta^K)$. Note that, for large $K$, this is significantly smaller compared to the complexity of computing APPs in~\cite{Lenz-itw20} (which is at least $\Omega(N K|\mathcal Q|\Delta^K u^K)$).
\begin{figure*}
\centering
\captionsetup[subfigure]{margin=10pt}
\subcaptionbox{TR error rates for unocded real DNA traces.\label{fig:real_uncoded_HRs}}
{\scalebox{0.7}{
\begin{tikzpicture}

\definecolor{color0}{rgb}{0.12156862745098,0.466666666666667,0.705882352941177}
\definecolor{color1}{rgb}{1,0.498039215686275,0.0549019607843137}
\definecolor{color2}{rgb}{0.172549019607843,0.627450980392157,0.172549019607843}

\begin{axis}[
legend cell align={left},
legend style={at={(0.975,0.61)},fill opacity=0.8, draw opacity=1, text opacity=1, draw=white!80!black},
tick align=outside,
tick pos=left,
x grid style={white!69.0196078431373!black},
xlabel={Number of traces},
xmajorgrids,
xmin=0.5, xmax=10.5,
xminorgrids,
xtick style={color=black},
y grid style={white!69.0196078431373!black},
ylabel={Normalized Hamming distance},
ymajorgrids,
ymin=0, ymax=0.4,
yminorgrids,
ytick style={color=black},
label style={font=\large},
]
\addplot [very thick, color0, mark=asterisk, mark size=3, mark options={solid}]
table {%
1 0.368
2 0.37077235624004
4 0.165160926030491
6 0.0860700796529673
8 0.0546306213304084
10 0.0422162404665769
};
\addlegendentry{BMALA}

\addplot [very thick, color1, mark=*, mark size=3, mark options={solid}]
table {%
1 0.3677
2 0.344950914200212
4 0.117595070336091
6 0.059573937648804
8 0.0394077037228208
10 0.0309093628070748
};
\addlegendentry{Trellis BMA}

\addplot [very thick, color2, mark=triangle*, mark size=3, mark options={solid,rotate=180}]
table {%
1 0.3677
2 0.334470761407735
4 0.290618171998134
6 0.271430886404148
8 0.260849087037873
10 0.254178392267221
};
\addlegendentry{Multiply posteriors}

\addplot [very thick, red, mark=square*, mark size=3, mark options={solid}]
table {%
1 0.3677
2 0.312213636363636
3 0.152177272727273
};
\addlegendentry{Multi-trace}

\end{axis}

\end{tikzpicture}}}
\subcaptionbox{TR error rates for real DNA traces with a rate-104/110 MR code.\label{fig:real_6coded_HRs}}
{\scalebox{0.7}{
\begin{tikzpicture}

\definecolor{color0}{rgb}{0.12156862745098,0.466666666666667,0.705882352941177}
\definecolor{color1}{rgb}{1,0.498039215686275,0.0549019607843137}
\definecolor{color2}{rgb}{0.172549019607843,0.627450980392157,0.172549019607843}

\begin{axis}[
legend cell align={left},
legend style={fill opacity=0.8, draw opacity=1, text opacity=1, draw=white!80!black},
tick align=outside,
tick pos=left,
x grid style={white!69.0196078431373!black},
xlabel={Number of traces},
xmajorgrids,
xmin=0.5, xmax=10.5,
xminorgrids,
xtick style={color=black},
y grid style={white!69.0196078431373!black},
ylabel={Normalized Hamming distance},
ymajorgrids,
ymin=0, ymax=0.3,
yminorgrids,
ytick style={color=black},
label style={font=\large}
]
\addplot [very thick, color0, mark=asterisk, mark size=3, mark options={solid}]
table {%
1 0.2321
2 0.2352353086496
4 0.0889001121751594
6 0.0415598730078368
8 0.0250382158650434
10 0.018780699635343
};
\addlegendentry{BMALA-MAP}
\addplot [very thick, color1, mark=*, mark size=3, mark options={solid}]
table {%
1 0.1854
2 0.104765617762406
4 0.0344211554041422
6 0.0187047115780147
8 0.0147813506578588
10 0.00971603915749272
};
\addlegendentry{Trellis BMA}
\addplot [very thick, color2, mark=triangle*, mark size=3, mark options={solid,rotate=180}]
table {%
1 0.1854
2 0.114004004158561
4 0.0571327302187416
6 0.0372222046430191
8 0.028133700933559
10 0.022890280797414
};
\addlegendentry{Multiply posteriors}
\end{axis}

\end{tikzpicture}}}
\subcaptionbox{TR error rates for real DNA traces with a rate-100/110 MR code.\label{fig:real_10coded_HRs}}
{\scalebox{0.7}{
\begin{tikzpicture}

\definecolor{color0}{rgb}{0.12156862745098,0.466666666666667,0.705882352941177}
\definecolor{color1}{rgb}{1,0.498039215686275,0.0549019607843137}
\definecolor{color2}{rgb}{0.172549019607843,0.627450980392157,0.172549019607843}

\begin{axis}[
legend cell align={left},
legend style={fill opacity=0.8, draw opacity=1, text opacity=1, draw=white!80!black, legend pos = north east},
tick align=outside,
tick pos=left,
x grid style={white!69.0196078431373!black},
xlabel={Number of traces},
xmajorgrids,
xmin=0.5, xmax=10.5,
xminorgrids,
xtick style={color=black},
y grid style={white!69.0196078431373!black},
ylabel={Normalized Hamming distance},
ymajorgrids,
ymin=0, ymax=0.3,
yminorgrids,
ytick style={color=black},
label style={font=\large},
]
\addplot [very thick, color0, mark=asterisk, mark size=3, mark options={solid}]
table {%
1 0.18807
2 0.190477719791249
4 0.0662908452605995
6 0.0289866995749349
8 0.0172902767920511
10 0.0123672797966203
};
\addlegendentry{BMALA-MAP}
\addplot [very thick, color1, mark=*, mark size=3, mark options={solid}]
table {%
1 0.14027
2 0.0673116552923859
4 0.0184012962462868
6 0.00885918003565062
8 0.0058736692689851
10 0.00395094960370869
};
\addlegendentry{Trellis BMA}
\addplot [very thick, color2, mark=triangle*, mark size=3, mark options={solid,rotate=180}]
table {%
1 0.14027
2 0.0729024488157367
4 0.0274966243586281
6 0.0149019607843137
8 0.00981121362668559
10 0.00734410049349484
};
\addlegendentry{Multiply posteriors}
\end{axis}

\end{tikzpicture}}}

\vspace{3mm}

\subcaptionbox{BCJR-once AIRs for TR of uncoded real DNA traces.\label{fig:real_uncoded_IRs}}
{\scalebox{0.7}{
\begin{tikzpicture}

\definecolor{color0}{rgb}{0.12156862745098,0.466666666666667,0.705882352941177}
\definecolor{color1}{rgb}{1,0.498039215686275,0.0549019607843137}
\definecolor{color2}{rgb}{0.172549019607843,0.627450980392157,0.172549019607843}

\begin{axis}[
legend cell align={left},
legend style={at={(0.975,0.59)},fill opacity=0.8, draw opacity=1, text opacity=1, draw=white!80!black},
tick align=outside,
tick pos=left,
x grid style={white!69.0196078431373!black},
xlabel={Number of traces},
xmajorgrids,
xmin=0.5, xmax=10.5,
xminorgrids,
xtick style={color=black},
y grid style={white!69.0196078431373!black},
ylabel={Rate (bits/base)},
ymajorgrids,
ymin=0.4, ymax=1.9,
ytick={0.6,1.0,1.4,1.8},
yminorgrids,
ytick style={color=black},
label style={font=\large}
]
\addplot [very thick, color0, mark=asterisk, mark size=3, mark options={solid}]
table {%
1 0.47
2 0.46
4 1.09
6 1.44
8 1.61
10 1.68
};
\addlegendentry{BMALA-HD}

\addplot [very thick, color1, mark=*, mark size=3, mark options={solid}]
table {%
1 0.688
2 0.828351986426516
4 1.25540025670444
6 1.47325406989732
8 1.53013687432891
10 1.52786986403473
};
\addlegendentry{Trellis BMA}
\addplot [very thick, color2, mark=triangle*, mark size=3, mark options={solid,rotate=180}]
table {%
1 0.688
2 0.872521175852658
4 0.839721652989365
6 0.700750802875711
8 0.494310244489044
10 0.310077837004101
};
\addlegendentry{Multiply posteriors}

\addplot [very thick, red, mark=square*, mark size=3, mark options={solid}]
table {%
1 0.688
2 0.997606620185793
3 1.37682536480773
};
\addlegendentry{ Multi-trace}

\end{axis}

\end{tikzpicture}}}
\subcaptionbox{BCJR-once AIRs for TR of real DNA traces with a rate-104/110 MR code.\label{fig:real_6coded_IRs}}
{\scalebox{0.7}{
\begin{tikzpicture}

\definecolor{color0}{rgb}{0.12156862745098,0.466666666666667,0.705882352941177}
\definecolor{color1}{rgb}{1,0.498039215686275,0.0549019607843137}
\definecolor{color2}{rgb}{0.172549019607843,0.627450980392157,0.172549019607843}

\begin{axis}[
legend cell align={left},
legend style={fill opacity=0.8, draw opacity=1, text opacity=1, legend pos=south east, draw=white!80!black},
tick align=outside,
tick pos=left,
x grid style={white!69.0196078431373!black},
xlabel={Number of traces},
xmajorgrids,
xmin=0.5, xmax=10.5,
xminorgrids,
xtick style={color=black},
y filter/.code={\pgfmathparse{(104*#1)/110}\pgfmathresult},
y grid style={white!69.0196078431373!black},
ylabel={Rate (bits/base)},
ymajorgrids,
ymin=0.4, ymax=1.9,
ytick={0.6,1.0,1.4,1.8},
yminorgrids,
ytick style={color=black},
label style={font=\large}
]
\addplot [very thick, color0, mark=asterisk, mark size=3, mark options={solid}]
table {%
1 1.034
2 1.02560937807718
4 1.59253552257584
6 1.73725914960535
8 1.81038509290391
10 1.82757300017414
};
\addlegendentry{BMALA-MAP}
\addplot [very thick, color1, mark=*, mark size=3, mark options={solid}]
table {%
1 1.236
2 1.61294682021868
4 1.87019548727652
6 1.91072799770152
8 1.94656298448909
10 1.95624338748515
};
\addlegendentry{Trellis BMA}
\addplot [very thick, color2, mark=triangle*, mark size=3, mark options={solid,rotate=180}]
table {%
1 1.236
2 1.58201097052485
4 1.77436990507153
6 1.82887150183533
8 1.86174509535017
10 1.87288333877831
};
\addlegendentry{Multiply posteriors}
\end{axis}

\end{tikzpicture}}}
\subcaptionbox{BCJR-once AIRs for TR of real DNA traces with a rate-100/110 MR code.\label{fig:real_10coded_IRs}}
{\scalebox{0.7}{
\begin{tikzpicture}

\definecolor{color0}{rgb}{0.12156862745098,0.466666666666667,0.705882352941177}
\definecolor{color1}{rgb}{1,0.498039215686275,0.0549019607843137}
\definecolor{color2}{rgb}{0.172549019607843,0.627450980392157,0.172549019607843}

\begin{axis}[
legend cell align={left},
legend style={fill opacity=0.8, draw opacity=1, text opacity=1, legend pos = south east, draw=white!80!black},
tick align=outside,
tick pos=left,
x grid style={white!69.0196078431373!black},
xlabel={Number of traces},
xmajorgrids,
xmin=0.5, xmax=10.5,
xminorgrids,
xtick style={color=black},
y filter/.code={\pgfmathparse{(100*#1)/110}\pgfmathresult},
ytick={0.6,1.0,1.4,1.8},
y grid style={white!69.0196078431373!black},
ylabel={Rate (bits/base)},
ymajorgrids,
ymin=0.4, ymax=1.9,
yminorgrids,
ytick style={color=black},
label style={font=\large}
]
\addplot [very thick, color0, mark=asterisk, mark size=3, mark options={solid}]
table {%
1 1.196
2 1.18280551907081
4 1.69413478765158
6 1.81770375560708
8 1.87394920421217
10 1.89180551032432
};
\addlegendentry{BMALA-MAP}
\addplot [very thick, color1, mark=*, mark size=3, mark options={solid}]
table {%
1 1.396
2 1.72964955099122
4 1.91881161081166
6 1.95343895019257
8 1.9752815897418
10 1.98177383043102
};
\addlegendentry{Trellis BMA}
\addplot [very thick, color2, mark=triangle*, mark size=3, mark options={solid,rotate=180}]
table {%
1 1.396
2 1.71103628150373
4 1.87931570191262
6 1.92774438803653
8 1.94774063092426
10 1.95656977547774
};
\addlegendentry{Multiply posteriors}
\end{axis}

\end{tikzpicture}}}
\caption{Experimental results on real data.  Note that Subfigures \ref{fig:real_6coded_IRs} and \ref{fig:real_10coded_IRs} include the rate loss of their MR codes.}\label{fig:numerics}
\vspace{-4mm}
\end{figure*}


\begin{figure*}
\centering
\captionsetup[subfigure]{margin=10pt}
\subcaptionbox{TR error rates for uncoded simulated DNA traces.\label{fig:sim_uncoded_HRs}}
{\scalebox{0.7}{
\begin{tikzpicture}

\definecolor{color0}{rgb}{0.12156862745098,0.466666666666667,0.705882352941177}
\definecolor{color1}{rgb}{1,0.498039215686275,0.0549019607843137}
\definecolor{color2}{rgb}{0.172549019607843,0.627450980392157,0.172549019607843}

\begin{axis}[
legend cell align={left},
legend style={at={(0.975,0.52)},fill opacity=0.8, draw opacity=1, text opacity=1, draw=white!80!black},
tick align=outside,
tick pos=left,
x grid style={white!69.0196078431373!black},
xlabel={Number of traces},
xmajorgrids,
xmin=0.5, xmax=10.5,
xminorgrids,
xtick style={color=black},
y grid style={white!69.0196078431373!black},
ylabel={Normalized Hamming distance},
ymajorgrids,
ymin=0, ymax=0.4,
yminorgrids,
ytick style={color=black},
label style={font=\large},
]
\addplot [very thick, color0, mark=asterisk, mark size=3, mark options={solid}]
table {%
1 0.371790303030303
2 0.374612121212121
4 0.120711515151515
6 0.0341454545454545
8 0.00573212121212121
10 0.000511515151515151
};
\addlegendentry{BMALA}

\addplot [very thick, color1, mark=*, mark size=3, mark options={solid}]
table {%
1 0.348210909090909
2 0.281673939393939
4 0.0459369696969697
6 0.0113151515151515
8 0.00202060606060606
10 0.000109090909090909
};
\addlegendentry{Trellis BMA}

\addplot [very thick, color2, mark=triangle*, mark size=3, mark options={solid,rotate=180}]
table {%
1 0.34832
2 0.306355151515152
4 0.258711515151515
6 0.244413333333333
8 0.237950303030303
10 0.220568484848485
};
\addlegendentry{Multiply posteriors}

\addplot [very thick, red, mark=square*, mark size=3, mark options={solid}]
table {%
1 0.343409090909091
2 0.2654
3 0.0616
};
\addlegendentry{Multi-trace}

\end{axis}

\end{tikzpicture}}}
\subcaptionbox{TR error rates for simulated DNA traces with a rate-104/110 MR code.\label{fig:sim_6coded_HRs}}
{\scalebox{0.7}{
\begin{tikzpicture}

\definecolor{color0}{rgb}{0.12156862745098,0.466666666666667,0.705882352941177}
\definecolor{color1}{rgb}{1,0.498039215686275,0.0549019607843137}
\definecolor{color2}{rgb}{0.172549019607843,0.627450980392157,0.172549019607843}

\begin{axis}[
legend cell align={left},
legend style={fill opacity=0.8, draw opacity=1, text opacity=1, draw=white!80!black, legend pos = north east},
tick align=outside,
tick pos=left,
x grid style={white!69.0196078431373!black},
xlabel={Number of traces},
xmajorgrids,
xmin=0.5, xmax=10.5,
xminorgrids,
xtick style={color=black},
y grid style={white!69.0196078431373!black},
ylabel={Normalized Hamming distance},
ymajorgrids,
ymin=0, ymax=0.3,
yminorgrids,
ytick style={color=black},
label style={font=\large},
]
\addplot [very thick, color0, mark=asterisk, mark size=3, mark options={solid}]
table {%
1 0.237501923076923
2 0.241086538461538
4 0.0654423076923077
6 0.0144153846153846
8 0.00319230769230769
10 0.00106730769230769
};
\addlegendentry{BMALA-MAP}
\addplot [very thick, color1, mark=*, mark size=3, mark options={solid}]
table {%
1 0.194055769230769
2 0.113328846153846
4 0.0199923076923077
6 0.00377884615384615
8 0.00155961538461538
10 0.000432692307692308
};
\addlegendentry{Trellis BMA}
\addplot [very thick, color2, mark=triangle*, mark size=3, mark options={solid,rotate=180}]
table {%
1 0.1938
2 0.123982692307692
4 0.0597807692307692
6 0.0371307692307692
8 0.0283461538461538
10 0.0216730769230769
};
\addlegendentry{Multiply posteriors}
\end{axis}

\end{tikzpicture}}}
\subcaptionbox{TR error rates for simulated DNA traces with a rate-100/110 MR code.\label{fig:sim_10coded_HRs}}
{\scalebox{0.7}{
\begin{tikzpicture}

\definecolor{color0}{rgb}{0.12156862745098,0.466666666666667,0.705882352941177}
\definecolor{color1}{rgb}{1,0.498039215686275,0.0549019607843137}
\definecolor{color2}{rgb}{0.172549019607843,0.627450980392157,0.172549019607843}

\begin{axis}[
legend cell align={left},
legend style={fill opacity=0.8, draw opacity=1, text opacity=1, draw=white!80!black, legend pos = north east},
tick align=outside,
tick pos=left,
x grid style={white!69.0196078431373!black},
xlabel={Number of traces},
xmajorgrids,
xmin=0.5, xmax=10.5,
xminorgrids,
xtick style={color=black},
y grid style={white!69.0196078431373!black},
ylabel={Normalized Hamming distance},
ymajorgrids,
ymin=0, ymax=0.3,
yminorgrids,
ytick style={color=black},
label style={font=\large},
]
\addplot [very thick, color0, mark=asterisk, mark size=3, mark options={solid}]
table {%
1 0.187538
2 0.19013
4 0.045354
6 0.008488
8 0.0024
10 0.0009
};
\addlegendentry{BMALA-MAP}
\addplot [very thick, color1, mark=*, mark size=3, mark options={solid}]
table {%
1 0.14826
2 0.071842
4 0.010588
6 0.00186
8 0.00015
10 0
};
\addlegendentry{Trellis BMA}
\addplot [very thick, color2, mark=triangle*, mark size=3, mark options={solid,rotate=180}]
table {%
1 0.148046
2 0.079078
4 0.027168
6 0.013392
8 0.007642
10 0.005984
};
\addlegendentry{Multiply posteriors}
\end{axis}

\end{tikzpicture}}}

\vspace{3mm}

\subcaptionbox{BCJR-once AIRs for TR of uncoded simulated DNA traces.\label{fig:sim_uncoded_IRs}}
{\scalebox{0.7}{
\begin{tikzpicture}

\definecolor{color0}{rgb}{0.12156862745098,0.466666666666667,0.705882352941177}
\definecolor{color1}{rgb}{1,0.498039215686275,0.0549019607843137}
\definecolor{color2}{rgb}{0.172549019607843,0.627450980392157,0.172549019607843}

\begin{axis}[
legend cell align={left},
legend style={at={(0.975,0.67)},fill opacity=0.8, draw opacity=1, text opacity=1, draw=white!80!black},
tick align=outside,
tick pos=left,
x grid style={white!69.0196078431373!black},
xlabel={Number of traces},
xmajorgrids,
xmin=0.5, xmax=10.5,
xminorgrids,
xtick style={color=black},
y grid style={white!69.0196078431373!black},
ylabel={Rate (bits/base)},
ymajorgrids,
ymin=0.4, ymax=2.05,
ytick={0.4,0.8,1.2,1.6,2.0},
yminorgrids,
ytick style={color=black},
label style={font=\large}
]
\addplot [very thick, color0, mark=asterisk, mark size=3, mark options={solid}]
table {%
1 0.46
2 0.45
4 1.28
6 1.73
8 1.94
10 1.99
};
\addlegendentry{BMALA-HD}

\addplot [very thick, color1, mark=*, mark size=3, mark options={solid}]
table {%
1 0.813125011997673
2 1.08760787705579
4 1.66342784413178
6 1.88543389960336
8 1.97067348110138
10 1.99804050734255
};
\addlegendentry{Trellis BMA}
\addplot [very thick, color2, mark=triangle*, mark size=3, mark options={solid,rotate=180}]
table {%
1 0.813180002860184
2 0.996991396075528
4 1.01833832852838
6 0.912893716423115
8 0.743878743982369
10 0.680531005066446
};
\addlegendentry{Multiply posteriors}

\addplot [very thick, red, mark=square*, mark size=3, mark options={solid}]
table {%
1 0.820842581802758
2 1.2122
3 1.76497
};
\addlegendentry{ Multi-trace}

\end{axis}

\end{tikzpicture}}}
\subcaptionbox{BCJR-once AIRs for TR of simulated DNA traces with a rate-104/110 MR code.\label{fig:sim_6coded_IRs}}
{\scalebox{0.7}{
\begin{tikzpicture}

\definecolor{color0}{rgb}{0.12156862745098,0.466666666666667,0.705882352941177}
\definecolor{color1}{rgb}{1,0.498039215686275,0.0549019607843137}
\definecolor{color2}{rgb}{0.172549019607843,0.627450980392157,0.172549019607843}

\begin{axis}[
legend cell align={left},
legend style={fill opacity=0.8, draw opacity=1, text opacity=1, legend pos = south east, draw=white!80!black},
tick align=outside,
tick pos=left,
x grid style={white!69.0196078431373!black},
xlabel={Number of traces},
xmajorgrids,
xmin=0.5, xmax=10.5,
xminorgrids,
xtick style={color=black},
y grid style={white!69.0196078431373!black},
ylabel={Rate (bits/base)},
ymajorgrids,
ymin=0.4, ymax=2.05,
ytick={0.4,0.8,1.2,1.6,2.0},
yminorgrids,
ytick style={color=black},
label style={font=\large}
]
\addplot [very thick, color0, mark=asterisk, mark size=3, mark options={solid}]
table {%
1 0.98510528
2 0.98227951
4 1.56840487
6 1.74357042
8 1.77568463
10 1.78474276
};

\addlegendentry{BMALA-MAP}
\addplot [very thick, color1, mark=*, mark size=3, mark options={solid}]
table {%
1 1.16436354
2 1.47817849
4 1.77917506
6 1.85007514
8  1.88175663
10 1.88028153
};
\addlegendentry{Trellis BMA}
\addplot [very thick, color2, mark=triangle*, mark size=3, mark options={solid,rotate=180}]
table {%
1 1.16453284
2 1.45639005
4 1.66158762
6 1.73374722
8  1.76365396
10 1.78997713
};
\addlegendentry{Multiply posteriors}
\end{axis}

\end{tikzpicture}}}
\subcaptionbox{BCJR-once AIRs for TR of simulated DNA traces with a rate-100/110 MR code.\label{fig:sim_10coded_IRs}}
{\scalebox{0.7}{
\begin{tikzpicture}

\definecolor{color0}{rgb}{0.12156862745098,0.466666666666667,0.705882352941177}
\definecolor{color1}{rgb}{1,0.498039215686275,0.0549019607843137}
\definecolor{color2}{rgb}{0.172549019607843,0.627450980392157,0.172549019607843}

\begin{axis}[
legend cell align={left},
legend style={fill opacity=0.8, draw opacity=1, text opacity=1, legend pos = south east, draw=white!80!black},
tick align=outside,
tick pos=left,
x grid style={white!69.0196078431373!black},
xlabel={Number of traces},
xmajorgrids,
xmin=0.5, xmax=10.5,
xminorgrids,
xtick style={color=black},
y grid style={white!69.0196078431373!black},
ylabel={Rate (bits/base)},
ymajorgrids,
ymin=0.4, ymax=2.05,
ytick={0.4,0.8,1.2,1.6,2.0},
yminorgrids,
ytick style={color=black},
label style={font=\large}
]
\addplot [very thick, color0, mark=asterisk, mark size=3, mark options={solid}]
table {%
1 1.10759604
2 1.09867739
4 1.60253694
6 1.73049614
8 1.75259655
10 1.75790698
};

\addlegendentry{BMALA-MAP}
\addplot [very thick, color1, mark=*, mark size=3, mark options={solid}]
table {%
1 1.24740314
2 1.56581543
4 1.75718551
6 1.80261226
8 1.81764501
10 1.81814968
};

\addlegendentry{Trellis BMA}
\addplot [very thick, color2, mark=triangle*, mark size=3, mark options={solid,rotate=180}]
table {%
1 1.24782158
2 1.54512574
4 1.71736223
6 1.76495811
8 1.78821048
10 1.79249098
};

\addlegendentry{Multiply posteriors}
\end{axis}

\end{tikzpicture}}}
\caption{Experimental results on simulated data.  Note that Subfigures \ref{fig:sim_6coded_IRs} and \ref{fig:sim_10coded_IRs} include the rate loss of their MR codes. These results are based on simulated data and are not affected by the issue discussed in “Note added on 8/12/2024” in Section III.}\label{fig:numerics_simulated}
\vspace{-4mm}
\end{figure*}

\subsection{Trellis BMA}
\label{sec:tbma}

Given the exponential growth of the multi-trace IDS trellis with the number of traces, we next describe a low-complexity heuristic that combines IDS trellises for individual traces to sequentially construct approximate posterior estimates,  $\U = (\widehat{U}_1, \widehat{U}_2, \ldots, \widehat{U}_L )$, for each message symbol.
This can be used to construct a hard estimate $\widehat{\M} = \widehat M_1 \widehat M_2...\widehat M_L$ for the message.

\paragraph{Initialization} Following the steps outlined in the previous subsection, we first construct $K$ independent trellises: one for each trace  $\y^k$ with $k\in [K]$.  Then, we run BCJR inference on each of the $K$ trellises with the corresponding traces as observations and compute $F^k(v)$ and $B^k(v)$, the forward and backward values of each vertex $v$ in the trellis corresponding to trace $k$, for all $k$ -- these values will be updated using a consensus across traces.
\gnote{Can we include the formulas for $F^k(v)$ and $B^k(v)$ in terms of what probabilities they represent?}
\hnote{Yes, in the long version}
\paragraph{Decoding} We now compute $\widehat \U$ by iterating through the following two steps. Working inductively, we assume that we have already computed $\widehat U_1,\widehat U_2...\widehat U_{l-1}$ and we would like to compute $\widehat U_l$.

$\bullet$ {\bf Combining beliefs from each trellis.} First, we use the current values of $F^k(v)$ and $B^k(v)$ to compute a ``belief'' about symbol $M_l$ for each trellis, denoted by $V^k(M_l=m)$. Recall that each $M_l$ is part of the trellis state in some stages (e.g., stages corresponding to input cycle $l$). Then, pick one of these (e.g., the last stage), call it stage $t$, and define \vspace{-0.0mm} \begin{equation} \label{eq:fb_output}
V^k(M_l=m) \triangleq \sum_{v \in \mathcal{V}_t (M_l = m)} F^k(v) \big( B^k(v) \big)^{\bb}, \vspace*{-0.5mm}
\end{equation}
where the sum is over stage-$t$ vertices with  on-deck message symbol $M_l=m$ and $\bb \geq 0$ reweights the backward values. 

\hspace{3mm} 
The channel outputs are conditionally independent given $\M$, so we have $\Pr(\Y^1,\Y^2,...,\Y^K|\M) = \prod_k \Pr(\Y^k|\M)$.
The RHS likelihoods can theoretically be combined to compute the true posterior.
However, BCJR inference outputs the marginals and multiplying them only gives the  approximation \vspace*{-1mm}
$$ V(M_l = m) \triangleq \prod\nolimits_{k=1}^K V^k(M_l=m). \vspace*{-0.0mm} $$

$\bullet$ {\bf Updating the forward values.}
For trellis $k$, the idea is to combine information from the other trellises to help maintain the correct synchronization on this trellis.
To do this, the forward BCJR values in stage $t$ are updated using the rule
$$F^k(v)\leftarrow \gamma^k \big( m(v) \big) F^k(v),$$
where $m(v)$ is value of $M_l$ associated with vertex $v$ and $\gamma^k (m)$ acts as a ``new prior'' for $M_l$ in trellis $k$ due to the other trellises.
We also note that the sum $\sum_m \gamma^k (m)$ does not affect the answer and, thus, $\gamma^k (\cdot)$ acts as an unnormalized probability.

\hspace{3mm} To define $\gamma^k (\cdot)$, we use the parametrized expression \vspace*{-0.5mm} $$\gamma^k (m) \triangleq (V^k(M_l=m))^{\bi}\prod\nolimits_{j\neq k}^K (V^j(M_l=m))^{\be}. \vspace*{-0.5mm}$$
This is motivated by the idea of extrinsic information processing~\cite{Berrou-icc93,Kschischang-it01}.
The parameter $\be \geq 0$ controls the dependence induced between the separate strand detectors, while $\bi \geq 0$ controls the intrinsic bias in each strand.
While $\be=1$  is a natural choice, smaller values of $\be$ reduce the dependence between strands and larger values push the $\gamma^k (\cdot)$ distribution towards a hard decision. Similarly, $\bi=0$  is a natural choice but larger values can sometimes improve performance.

\hspace{3mm} For the posterior estimate of $M_l$ given $\Y^1,\Y^2,...,\Y^K$, we define
$ \widehat{U}_{l} (m) \triangleq c_l V(M_l = m)^\bo$
for some $\bo>0$ and choose $c_l$ so the sum over $m$ equals 1.
To lower bound the AIR, we apply the RHS of~\eqref{eq:CodedTR_smap} to $\widehat{U}_l$.
Choosing $\bo<1$ may mitigate overconfidence and increase the AIR lower bound.
 

\hspace{3mm} Using the updated forward values at input cycle $l$, we then continue the forward pass to input cycle $l+1$ and compute $V^k (M_l = m)$.
Then, this is used to update the forward values for the vertices of input cycle $l+1$.
This process repeats for the first half of the inputs.

\paragraph{Estimating each half} Using this updating approach, we sequentially compute the estimates  $\widehat U_1 \widehat U_2...\widehat U_{L/2}$. Analogously, we start from the end of the trellis and update the backward values to compute an estimate $\widehat U_{L/2+1} \widehat U_2...\widehat U_L$ which proceeds in the reverse order.
For the reverse estimate, \eqref{eq:fb_output} should use the first stage with $U_l$ in the state.

\paragraph{Time Complexity} 
The time complexity is $K$ times the complexity of computing APPs using the multi-trace trellis with one trace, which is equal to $O(KN|\mathcal Q|\Delta)$.

\vspace{-0.25mm}
\section{Experimental results}
\vspace{-0.25mm}
\label{sec:numerics}
In Fig.~\ref{fig:numerics}, we provide experimental results, with and without coding, for the algorithm introduced in this paper.
We also compare to previous approaches such as ``separate decoding'' using ``multiply posteriors'' from \cite{Lenz-itw20}, BMALA (see Appendix~\ref{app:BMALA}) from \cite[pp.~6--7]{DNA-assembly2019}\cite{BMA-20}, and to BCJR on the multi-trace IDS trellis from Section~\ref{subset:multiD} (see also~\cite{Lenz-itw20}).
Note that BMALA is a TR algorithm and does not give soft output. Hence, the BMALA-HD curve in Fig.~\ref{fig:numerics}(d) maps the hard-decision symbol error rate into an achievable rate.
We note that BMALA-HD beats Trellis BMA for more than 6 traces even though Trellis BMA has a lower error rate.
This is because the soft outputs of Trellis BMA are not ideally calibrated.
In future work, we will investigate learning-based methods to see if they can generate better calibrated output probabilities.

For coded TR, we use BMALA to give a hard estimate of the DNA sequence and treat this estimate as an observed trace for IDS trellis decoding of the message symbols; we call this BMALA-MAP. We also report the numbers for  the multi-trace trellis only for TR with 3 or fewer; other experiments with the multi-trace trellis are computationally infeasible.

The 10000 clusters of DNA sequences (and corresponding traces) in the datset are divided  into training (clusters 1-2000), validation (clusters 2001-2500) and test sets (clusters 2501-10000). Training is used to learn the IDS channel ($\pins$, $\pdel$, $\psub$), validation is used to tune the hyperparameters ($\beta_e, \beta_o$, etc.) for Trellis BMA, and the test set is used for the reported results. We remark that multiply posteriors is an instance of Trellis BMA when $\beta_b=\beta_o=1$ and $\beta_e=\beta_i=0$.

\vspace{-0.25mm}
\section*{Acknowledgment}
\vspace{-0.25mm}

We thank Karin Strauss, Yuan-Jyue Chen, and the Molecular Information Systems Laboratory (MISL) for providing the DNA dataset released with this paper and useful discussions on this topic.

\bibliographystyle{IEEEtran}
\bibliography{main}


\appendix

\subsection{BMALA for IDS channel}
\label{app:BMALA}
Consider the  DNA storage architecture, shown in Figure~\ref{fig:dna_storage} and described in \cite{Organick-natbiotech18}.
Ignoring the outer code, this scheme uses an identity map to encode the message sequence into a DNA sequence.
For the decoder, it uses the Improved BMALA TR algorithm to estimate
the DNA sequence.

Now, wee briefly describe the steps in the Improved BMALA algorithm, which attempts to sequentially estimate each symbol of the DNA sequene $\X$.
For each of the traces, it uses a hard estimate of the input pointer and then estimates the 
next symbol of $\X$ using a vote of the \emph{current} symbols implied by the hard input-pointer estimates.  For the traces that do not agree with the plurality, it tries to infer the
reason for disagreement (e.g., insertion, deletion, or substitution) by looking ahead 
a few symbols, and moves
the corresponding pointers accordingly. If the algorithm cannot decide on any reason for disagreement,
it discards the trace temporarily and attempts to brings it back at a later point in time.
\hnote{Better reference?}

\subsection{Random scrambling induces uniform distribution}
\label{app:random_scrabling}

Suppose $\Sigma$ has an abelian group structure, and let code $\mathcal{C}\subseteq \Sigma^N$. Let $\mathbf C \sim\ \mathrm{Uniform}(\mathcal C)$, $\mathbf Z\sim\ \mathrm{Uniform}(\Sigma^N)$ and $\mathbf X = \mathbf C + \mathbf Z$, where $+$ is defined as the coordinate-wise application of the group operation. We show here that $\mathbf X$ is uniformly distributed on $\Sigma^N$. We do so by proving the following stronger claim
$$\Pr(\mathbf X = \mathbf x|\mathbf C = \mathbf c)=\frac{1}{|\Sigma|^N},\ \forall\ \mathbf x \in \Sigma^N, \mathbf c \in \mathcal C.$$ As a result $$\Pr(\mathbf X=\mathbf x)= \mathbb E_{\mathbf C} \Pr(\mathbf X = \mathbf x|\mathbf C) = \frac{1}{|\Sigma|^N},\ \forall\ \mathbf x \in \Sigma^N.$$

Before proving our claim we first define $\mathbf v^{-1} \in \Sigma^N$ to be the coordinate-wise inverse of $\mathbf v \in \Sigma^N$. Since $\Sigma$ has a group structure, $\mathbf v^{-1}$ exists and is unique for every $\mathbf v \in \Sigma^N$. To prove our claim, we observe that 
\begin{align*}
    \Pr(\mathbf X = \mathbf x|\mathbf C = \mathbf c) &= \Pr(\mathbf Z + \mathbf c=\mathbf x)\\
    & = \Pr(\mathbf Z + \mathbf c + \mathbf c^{-1}=\mathbf x + \mathbf c^{-1})\\
    & = \Pr(\mathbf Z =\mathbf x + \mathbf c^{-1})\\
    & = \frac{1}{|\Sigma|^N},
\end{align*}
which concludes the proof.

\subsection{Trellis structure and BCJR inference}
\label{app:prob_fsm}
In this section, we outline the essential tools used in this work.  Crucially, we describe our \textit{trellis} structure. We remark that our trellis definition differs somewhat from standard definitions used in the coding theory literature. This variation is essential to efficiently represent a larger class
of input-output distributions, such as the one that describes the IDS channel. In most standard applications, the states in a trellis are organized into distinct stages and edges may only connect states in adjacent stages. While it is possible to represent IDS channels in this fashion, it requires many more edges.

Our trellis is a directed acyclic graph (DAG) that describes the joint distribution for a collection of observed random sequences $\Y^1,\Y^2,...,\Y^K$ and a latent or hidden sequence of states.
The state sequence is $\S = (S_1,S_2,...,S_L)$, where the length $L$ is a random variable satisfying $L\leq c$ for some constant $c$. We assume that $S_1$ alone is known apriori and fixed to be $s_0$. The support of each symbol in $\Y^k$ is a finite set $\mathcal Y$. Likewise, the support of $S_i$ is a finite set $\mathcal S$.
Let $\y^k = y^k_1y^k_2...,y^k_{R_k}$ denote the observed realization of $\Y^k$, where $R_k$ is the length of the $k$-th observable sequence and $s=(s_1,s_2,\ldots,s_l)$ denote a possible realization of $S$.
The trellis describes the joint distribution $$\Pr(\S=\s,\Y^1=\y^1,\Y^2=\y^2,...,\Y^K=\y^K).$$ We now describe essential properties of the trellis DAG, and define some useful notation.

\begin{itemize}[leftmargin=2mm]
\setlength\itemsep{0mm} 
\item \textbf{Vertices:} The vertices in the trellis are all possible state realizations. Each vertex is uniquely identified by a state $s \in \mathcal S$. The trellis has exactly one origin  $s_0$ (state with no in-neighbors) and a set of absorbing states $\mathcal S_{abs}$ (states with no out-neighbors).
\item \textbf{Edges:} Suppose edge $e$ connects vertex  $s$ to vertex $s'$. Define $\head(e) = s$ as the from vertex of $e$, $\tail(e)= s'$ as the to vertex of $e$. 

 \item{\bf Edge labels:} An edge can either have no label (\textit{unlabeled edge}), or is labeled by the (trace, symbol) pair $(k,j)$, corresponding to the observation $y^k_j$; this edge is one explanation of the observed symbol $y^k_j$.  Multiple edges can have the same label. Define $\elabel(e) = (k,j)$ as the label of $e$. For an unlabeled edge $\elabel(e) = \phi$.
 
\item \textbf{Edge weights:} Every edge in the trellis is weighted. The weight of an unlabeled edge connecting vertices $s$ and $s'$ is equal to $\Pr(s'|s)$. For an edge with label $(k,j)$ that connects the vertices $s$ and $s'$, the edge weight is equal to $\Pr(\Y^k_j=y^k_j,s'|s)$.
For a vertex which is not an absorbing state, the weights of all its outgoing edges should sum to 1.

\item \textbf{Paths:} For an edge path $p=e_1e_2...e_L$, define $\head(p)\triangleq \head(e_1)$ and $\tail(p)\triangleq \tail(e_L)$. For the trellis to describe the joint distribution $\Pr(\S = \s,\Y^1=\y^1,\Y^2=\y^2,...,\Y^K=\y^K),$ the following property needs to be satisfied: consider a path $p = e_1e_2...e_L$ where $\head(p)$ is the origin and $\tail(p)$ is an absorbing state (for every $k,j$, there exists exactly one edge $e_l$ in every such path such that $\elabel(e_l)=(k,j)$). In other words, every path that connects the origin to an absorbing state must explain all the observed symbols exactly once.

\end{itemize}

\textbf{Remark.} One can verify that the IDS trellis described in section~\ref{subset:multiD} satisfies the above properties.

The weight $w(p)$ of a path $p$ is defined to be the product of weights of the constituent edges. Each path $p=e_1e_2...e_L$ in the trellis connecting the origin to an absorbing state corresponds to a particular sequence of states $\S=(s_0,\head(e_2),...,\head(e_L),\tail(e_L))$ with the given observations $\Y^1=\y^1,...,\Y^K=\y^K$. Moreover, path weight $w(p)$ of a path $p=e_1e_2...e_L$ is
\begin{align*}
w(p) &= \Pr(\S=(s_0,\tail(e_1),...,\tail(e_L))\\&\hspace{1cm},\Y^1=\y^1,...,\Y^K=\y^K|S_1=s_o)\\
&\overset{(a)}{=} \Pr(\S=(s_0,\tail(e_1),...,\tail(e_L))\\&\hspace{1cm},\Y^1=\y^1,...,\Y^K=\y^K),
\numberthis
\label{eq:path_wt}
\end{align*}
where $(a)$ follows since $\Pr(S_1=s_0)=1$, as $S_1$ is known and fixed to be $s_0$ apriori.

We next describe the forward-backward algorithm (also called the BCJR algorithm \cite{Bahl-it74}) that computes the probability that the hidden state $s$ was encountered during the output generation process~\cite{Davey-it01}.
Abusing notation, we denote this probability by
\begin{align*}
\Pr(&s\in \S,\Y^1=\y^1,...,\Y^K=\y^K) \\ &\triangleq
\sum_{\s : \exists i\in \{1,\ldots,|\s|\}, s_i = s} \Pr(\S = \s,\Y^1=\y^1,...,\Y^K=\y^K),
\end{align*}
where $|\s|$ represents the length of $\s$.
For marginal inference of the input symbols, it is sufficient to compute $\Pr(s \in \S,\Y^1=\y^1,...,\Y^K=\y^K)$
for all $s\in \mathcal{S}$
because the input symbols are deterministic functions of the state.

To compute this quantity, we interpret $\Pr(s \in \S,\Y^1=\y^1,...,\Y^K=\y^K)$ as the sum of the weights of all paths that start at the origin, end at an absorbing state and pass through state $s$ in the trellis.
Then, the derivation of BCJR inference reveals this probability as the product of two terms via the decomposiiton
\begin{align*}
\Pr&(s \in \S,\Y^1=\y^1,...,\Y^K=\y^K)=\sum_{\substack{p: \head(p)=s_0,\\ \tail(p)\in \mathcal S_{abs}}} w(p) \\
&\overset{(a)}{=} \sum_{\substack{p_1,p_2:\\ \head(p_1)=s_0, \tail(p_1)=s\\\head(p_2)=s, \tail(p_2)\in \mathcal S_{abs}}} w(p_1)w(p_2) \\
    &\hspace{0.5cm}= \left(\sum_{\substack{p_1: \head(p_1)=s_0,\\ \tail(p_1=s)}} w(p_1)\right)\left( \sum_{\substack{p_2:\head(p_2)=s,\\ \tail(p_2)\in \mathcal S_{abs}}} w(p_2)\right),
\end{align*}
where in $(a)$ we split each path $p$ into two paths such that the first path ends at $s$ and the second originates at $s$.

For each state $s\in \mathcal{S}$, we define the forward value to be $$F(s)\triangleq \sum_{\substack{p: \head(p)=s_0,\\ \tail(p)=s}} w(p)$$ and the backward value to be
$$B(s) \triangleq \sum_{\substack{p:\head(p)=s,\\ \tail(p)\in \mathcal S_{abs}}} w(p).$$
Together, these imply that 
$$\Pr(s \in \S,\Y^1=\y^1,...,\Y^K=\y^K)= F(s)B(s).$$

\paragraph{Computing the forward values for each state}
We now present the dynamic program that computes $F(s)$ for all $s$. But first some notation: define $\mathcal E_{in}(s)$ as the set of edges whose tail is $s$.

\begin{align*}
F(s) &= \sum_{\substack{p: \head(p)=s_0,\\ \tail(p)=s}} w(p),\\
&\overset{(a)}{=} \sum_{e \in \mathcal E_{in}(s)} \sum_{\substack{p': \head(p')=s_0,\\ \tail(p')=\head(e)}}  w(p'e)\\
&= \sum_{e \in \mathcal E_{in}(s)} \sum_{\substack{p': \head(p')=s_0,\\ \tail(p')=\head(e)}}  w(p')w(e) \\
&= \sum_{e \in \mathcal E_{in}(s)} w(e) \sum_{\substack{p': \head(p')=s_0,\\ \tail(p')=\head(e)}} w(p') \\
&= \sum_{e \in \mathcal E_{in}(s)} w(e) F(\head(e)), \numberthis
\label{eq:forward_pass}
\end{align*}
where in $(a)$, we split the path $p$ as $p'e$, where $\tail(e)=s$.

 To compute the forward values of all vertices, we first compute a topological ordering for the vertices of the trellis. Recall that the trellis is a DAG, so such an ordering always exists (see \cite{knuth1974structured} and references therein). Next we initialize the forward values $F(s_0)=1$. Since all paths begin there, this is sufficient.
 Next, we traverse the vertices in order and use the aforementioned sum-product update rule in \eqref{eq:forward_pass} to compute $F(s)$ for all vertices in the trellis. Since each edge in the trellis is traversed exactly once when computing the forward values, the complexity of computing the forward values is $O(E),$ where $E$ is the number of edges in the trellis. Moreover, a topological ordering (done once offline) can be accomplished by a bread-first search (whose complexity is $O(E)$ as well) starting from the origin state, and hence does not affect the overall complexity of our algorithm.

\paragraph{Computing the backward values for each state}
We next present the dynamic program that computes $B(s)$ for all $s$. But first some notation: define $\mathcal E_{out}(s)$ as the set of edges whose head is $s$.

\begin{align*}
B(s) &= \sum_{\substack{p: \head(p)=s,\\ \tail(p)\in \mathcal S_{abs}}} w(p),\\
&\overset{(a)}{=} \sum_{e \in \mathcal E_{out}(s)} \sum_{\substack{p': \head(p')=\tail(e),\\ \tail(p')\in \mathcal S_{abs}}}  w(ep')\\
&= \sum_{e \in \mathcal E_{out}(s)} \sum_{\substack{p': \head(p')=\tail(e),\\ \tail(p')\in \mathcal S_{abs}}}  w(p')w(e) \\
&= \sum_{e \in \mathcal E_{out}(s)} w(e) \sum_{\substack{p': \head(p')=\tail(e),\\ \tail(p')\in \mathcal S_{abs}}} w(p') \\
&= \sum_{e \in \mathcal E_{out}(s)} w(e) B(\tail(e)), \numberthis
\label{eq:backward_pass}
\end{align*}
where in $(a)$, we split the path $p$ as $ep'$, where $\head(e)=s$.

To compute the backward values of all vertices, we use the reverse topological ordering for the vertices of the trellis. Next we initialize the backward values of the abosrbing states $B(s)=1\ \forall s\in \mathcal S_{abs}.$ The complexity of computing the backward values is also $O(E),$ since each edge is traversed exactly once.

\paragraph{Output stage}
Recall that in the IDS trellis, the states (vertices) are of the form $(q_t, p^1_t, p^2_t,...,p^K_t, m_t, x_t)$ where $t$ designates the stage, $q_t$ is the state of the encoder, $p^k_t$ are the pointer values, $m_t$ is the message symbol and $x_t$ is the codeword symbol. Therefore, the message symbol is itself a part of the state. To compute the posterior distribution of the $i$-th message symbol $M_i$, we first compute $\Pr(s\in \S,\Y^1=\y^1,...,\Y^K=\y^K)$ for all vertices $s$ in the trellis.

Recall that there are multiple stages in the trellis for each input (e.g., the input stage and the stages associated with the outputs for each of the $K$ traces).
To compute the output, we focus on the last stage in the trellis associated with input $i$ which is right before transitioning to message $i+1$.
This stage has no intra-stage edges.
For each $m\in \mathcal{M}$, we define $\mathcal{S}_m$ to be the subset of states in this stage associated with $M_i = m$ and compute
$$V (M_i = m) = c_i \sum_{s \in \mathcal{S}_m} \Pr(s\in \S,\Y^1=\y^1,...,\Y^K=\y^K),$$
where $c_i$ is chosen so that $\sum_{m\in \mathcal{M}} V (M_i = m) = 1$.
Then,
\[V (M_i = m) = \Pr(M_i = m | \Y^1=\y^1,...,\Y^K=\y^K). \]

\paragraph{Complexity analysis} The forward and backward passes traverse the trellis edges exactly once. Moreover, finding a topological order for the trellis vertices is $O(E)$ and this is done once offline. The number of vertices is at most twice the number of edges. The time complexity of forward-backward algorithm is $O(E)$.

\subsection{Convolutional codes (CC) vs. Marker repeat (MR) codes}
\label{app:MR-vs-CC}
In Fig.~\ref{fig:CC-vs-MR-HR} and Fig.~\ref{fig:CC-vs-MR-IR}, we compare the relative performance of CC and MR  for a few different coding rates using the dataset and approach from Section~\ref{sec:dataset}.  The idea is to investigate which choice of code is appropriate given a fixed inner coding rate. For illustration purposes, we fix the number of observed traces to 2 and use the following set of $\beta$ values to decode via Trellis BMA -- $\beta_b=1, \beta_e = 0.1, \beta_i = 0, \beta_o = 1.0$. We observed similar performance with other sets of $\beta$ values and we strongly suspect that the relative performance of CC and MR codes is insensitive to the particular choice of $\beta$s.

The following plots illustrate that MR codes clearly outperform CC when the inner coding rate is 0.9 or more. For lower rates of inner codes, the MR codes are only marginally worse than CC. Moreover, the time taken to decode with MR codes is also lower, since the IDS trellis constructed with CC has a larger state-space. 


\begin{figure}[!h]
    \centering
    \scalebox{0.8}{
\begin{tikzpicture}
\definecolor{color0}{rgb}{0.12156862745098,0.466666666666667,0.705882352941177}
\definecolor{color1}{rgb}{1,0.498039215686275,0.0549019607843137}
\definecolor{color2}{rgb}{0.172549019607843,0.627450980392157,0.172549019607843}

\begin{axis}[
legend cell align={left},
legend pos = north west,
legend style={fill opacity=0.8, draw opacity=1, text opacity=1, draw=white!80!black},
tick align=outside,
tick pos=left,
x grid style={white!69.0196078431373!black},
xlabel={Rate of inner code},
xmajorgrids,
xmin=0.5, xmax=1,
xminorgrids,
xtick style={color=black},
y grid style={white!69.0196078431373!black},
ylabel={Normalized Hamming distance},
ymajorgrids,
ymin=0, ymax=0.35,
yminorgrids,
ytick style={color=black},
label style={font=\large},
]
\addplot [very thick, color0, mark=asterisk, mark size=3, mark options={solid}]
table {%
0.5454 0.012283333333333334
0.7272 0.022875
0.9090 0.067311
};
\addlegendentry{MR code}

\addplot [very thick, color1, mark=asterisk, mark size=3, mark options={solid}]
table {%
0.5454 0.00472
0.7272 0.018375
0.9090 0.30728
};
\addlegendentry{CC code}
\end{axis}

\end{tikzpicture}}
    \caption{Hamming error rate for convolutional codes (CC) and marker repeat (MR) codes evaluated using real data for different coding rates with 2 traces.}
    \label{fig:CC-vs-MR-HR}
\end{figure}

\begin{figure}[!h]
    \centering
    \scalebox{0.8}{
\begin{tikzpicture}
\definecolor{color0}{rgb}{0.12156862745098,0.466666666666667,0.705882352941177}
\definecolor{color1}{rgb}{1,0.498039215686275,0.0549019607843137}
\definecolor{color2}{rgb}{0.172549019607843,0.627450980392157,0.172549019607843}

\begin{axis}[
legend cell align={left},
legend pos = south west,
legend style={fill opacity=0.8, draw opacity=1, text opacity=1, draw=white!80!black},
tick align=outside,
tick pos=left,
x grid style={white!69.0196078431373!black},
xlabel={Rate of inner code},
xmajorgrids,
xmin=0.5, xmax=1,
xminorgrids,
xtick style={color=black},
y grid style={white!69.0196078431373!black},
ylabel={Rate (in bits/base)},
ymajorgrids,
ymin=0.6, ymax=1.8,
yminorgrids,
ytick style={color=black},
label style={font=\large},
]
\addplot [very thick, color0, mark=asterisk, mark size=3, mark options={solid}]
table {
0.5454 1.0602576
0.7272 1.38168
0.9090 1.57257
};
\addlegendentry{MR code}

\addplot [very thick, color1, mark=asterisk, mark size=3, mark options={solid}]
table {
0.5454 1.0788012
0.7272 1.396224
0.9090 0.887184
};
\addlegendentry{CC code}
\end{axis}

\end{tikzpicture}}
    \caption{AIRs for convolutional codes (CC) and marker repeat (MR) codes evaluated using real data for different coding rates with 2 traces. The rate loss of the inner code is included in these AIRs.}
    \label{fig:CC-vs-MR-IR}
\end{figure}


\subsection{Visual example of an IDS trellis}
\begin{figure*}
\centering
\includegraphics[width=1.9\columnwidth]{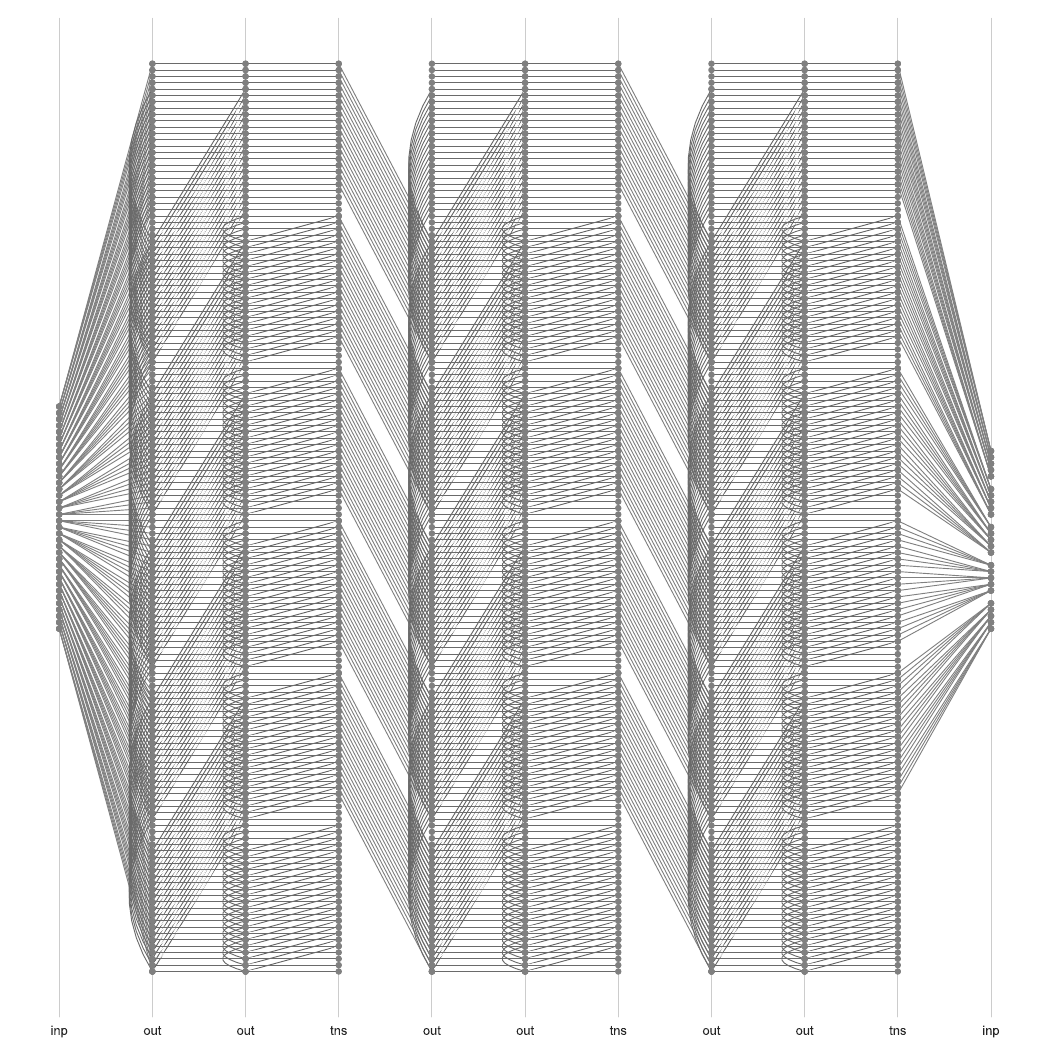}
\caption{\small \textbf{The multi-trace IDS trellis.} Example of 1 input cycle in the multi-trace IDS trellis for 2 traces and rate 1/3 encoder. Intra-stage edges are drawn as curved lines. The arrows on the directional edges have been removed to declutter the graph and for aesthetics. The first stage models the input and appends the first codeword to the output buffer, next models all possible events with the first codeword symbol in the first trace, then models events in the second trace. Next, it replaces the codeword symbol in the output buffer and models the IDS events with the second codeword symbol in the two traces. Finally it models the IDS events with the third codeword symbol in the two traces before transitioning to the next input symbol.}
\label{fig:trellis_2traces_code}
\end{figure*}

Please see Fig.~\ref{fig:trellis_2traces_code} for an example visualization of the IDS trellis. 


\subsection{Optimized hyperparameters for Trellis BMA on real data}
\label{app:betas}
\begin{center}
\begin{tabular}{ | m{1.5cm} || m{1cm} | m{1cm}| m{1cm} | m{1cm}| } 
\hline
\multicolumn{5}{|c|}{Hamming distance} \\
\hline
\hline
 Traces & $\beta_b$ &  $\beta_e$ & $\beta_i$ & $\beta_0$\\
 \hline
1 & 1  & 1.0 & 0 & 1.0\\
\hline
2 & 0  & 0.1 & 0.5 & 0.5\\ 
\hline
4 & 0 &  1 & 0.1 & 0.9\\ 
\hline
6 & 0 &  0.5 & 0.1 & 1\\ 
\hline
8 & 0 &  0.5 & 0.5 & 0.9\\ 
\hline
10 & 0 &  0.5 & 0 & 1\\ 
\hline
\end{tabular}
\end{center}

\begin{center}
\begin{tabular}{ | m{1.5cm} || m{1cm} | m{1cm}| m{1cm} | m{1cm}| } 
\hline
\multicolumn{5}{|c|}{BCJR-once rate} \\
\hline
\hline
 Traces & $\beta_b$ &  $\beta_e$ & $\beta_i$ & $\beta_0$\\
 \hline
1 & 1  & 1.0 & 0 & 1.0\\
\hline
2 & 0  & 0.05 & 0.5 & 0.5\\ 
\hline
4 & 0 &  0.5 & 0.1 & 0.5\\ 
\hline
6 & 0 &  0.5 & 0.1 & 0.5\\ 
\hline
8 & 0 &  0.5 & 0.5 & 0.5\\ 
\hline
10 & 0 &  1.0 & 0 & 0.5\\ 
\hline
\end{tabular}
\end{center}

\begin{center}
\begin{tabular}{ | m{1.5cm} || m{1cm} | m{1cm}| m{1cm} | m{1cm}| } 
\hline
\multicolumn{5}{|c|}{104/110 MR-coded Hamming distance} \\
\hline
\hline
 Traces & $\beta_b$ &  $\beta_e$ & $\beta_i$ & $\beta_0$\\
 \hline
1 & 1  & 1.0 & 0 & 1.0\\
\hline
2 & 1  & 0.1 & 0 & 0.5\\ 
\hline
4 & 1 &  0.1 & 0 & 0.5\\ 
\hline
6 & 1 &  0.1 & 0 & 0.5\\ 
\hline
8 & 1 &  0.1 & 0 & 0.5\\ 
\hline
10 & 1 &  0.02 & 0 & 0.5\\ 
\hline
\end{tabular}
\end{center}

\begin{center}
\begin{tabular}{ | m{1.5cm} || m{1cm} | m{1cm}| m{1cm} | m{1cm}| } 
\hline
\multicolumn{5}{|c|}{104/110 MR-coded BCJR-once rate} \\
\hline
\hline
 Traces & $\beta_b$ &  $\beta_e$ & $\beta_i$ & $\beta_0$\\
  \hline
1 & 1  & 1.0 & 0 & 1.0\\
\hline
2 & 1  & 0.1 & 0 & 1.0\\ 
\hline
4 & 1 &  0.1 & 0 & 0.5\\ 
\hline
6 & 1 &  0.02 & 0 & 0.5\\ 
\hline
8 & 1 &  0.02 & 0 & 0.5\\ 
\hline
10 & 1 &  0.02 & 0 & 0.5\\ 
\hline
\end{tabular}
\end{center}

\begin{center}
\begin{tabular}{ | m{1.5cm} || m{1cm} | m{1cm}| m{1cm} | m{1cm}| } 
\hline
\multicolumn{5}{|c|}{100/110 MR-coded Hamming distance} \\
\hline
\hline
 Traces & $\beta_b$ &  $\beta_e$ & $\beta_i$ & $\beta_0$\\
 \hline
1 & 1  & 1.0 & 0 & 1.0\\
\hline
2 & 1  & 0.1 & 0 & 0.1\\ 
\hline
4 & 1 &  0.1 & 0 & 0.1\\ 
\hline
6 & 1 &  0.1 & 0 & 0.1\\ 
\hline
8 & 1 &  0.02 & 0 & 1.0\\ 
\hline
10 & 1 &  0.02 & 0 & 1.0\\ 
\hline
\end{tabular}
\end{center}

\begin{center}
\begin{tabular}{ | m{1.5cm} || m{1cm} | m{1cm}| m{1cm} | m{1cm}| } 
\hline
\multicolumn{5}{|c|}{100/110 MR-coded BCJR-once rates} \\
\hline
\hline
 Traces & $\beta_b$ &  $\beta_e$ & $\beta_i$ & $\beta_0$\\
 \hline
1 & 1  & 1.0 & 0 & 1.0\\
\hline
2 & 1  & 0.1 & 0 & 1.0\\ 
\hline
4 & 1 &  0.1 & 0 & 0.5\\ 
\hline
6 & 1 &  0.1 & 0 & 0.5\\ 
\hline
8 & 1 &  0.02 & 0 & 0.5\\ 
\hline
10 & 1 &  0.02 & 0 & 0.5\\ 
\hline
\end{tabular}
\end{center}

\subsection{Optimized hyperparameters for Trellis BMA on simulated data}
\label{app:betas_sim}
\begin{center}
\begin{tabular}{ | m{1.5cm} || m{1cm} | m{1cm}| m{1cm} | m{1cm}| } 
\hline
\multicolumn{5}{|c|}{Hamming distance} \\
\hline
\hline
 Traces & $\beta_b$ &  $\beta_e$ & $\beta_i$ & $\beta_0$\\
 \hline
1 & 1  & 0.5 & 0 & 0.1\\
\hline
2 & 1  & 0.1 & 0.0 & 0.1\\ 
\hline
4 & 0 &  1.0 & 0.5 & 0.5\\ 
\hline
6 & 0 &  0.5 & 0.1 & 1.0\\ 
\hline
8 & 0 &  5.0 & 0.0 & 0.1\\ 
\hline
10 & 0 &  0.5 & 0.0 & 0.1\\ 
\hline
\end{tabular}
\end{center}

\begin{center}
\begin{tabular}{ | m{1.5cm} || m{1cm} | m{1cm}| m{1cm} | m{1cm}| } 
\hline
\multicolumn{5}{|c|}{BCJR-once rate} \\
\hline
\hline
 Traces & $\beta_b$ &  $\beta_e$ & $\beta_i$ & $\beta_0$\\
 \hline
1 & 1  & 0.5 & 0.0 & 1.0\\
\hline
2 & 1  & 0.1 & 0.0 & 0.5\\ 
\hline
4 & 0 &  1.0 & 0.5 & 0.5\\ 
\hline
6 & 0 &  0.5 & 0.1 & 0.5\\ 
\hline
8 & 0 &  5.0 & 0.0 & 0.5\\ 
\hline
10 & 0 &  0.5 & 0.1 & 1.0\\ 
\hline
\end{tabular}
\end{center}

\begin{center}
\begin{tabular}{ | m{1.5cm} || m{1cm} | m{1cm}| m{1cm} | m{1cm}| } 
\hline
\multicolumn{5}{|c|}{104/110 MR-coded Hamming distance} \\
\hline
\hline
 Traces & $\beta_b$ &  $\beta_e$ & $\beta_i$ & $\beta_0$\\
 \hline
1 & 1  & 1.0 & 0.0 & 0.5\\
\hline
2 & 1  & 0.1 & 0.0 & 0.1\\ 
\hline
4 & 1 &  0.5 & 0.5 & 1.0\\ 
\hline
6 & 1 &  5.0 & 0.1 & 0.5\\ 
\hline
8 & 1 &  1.0 & 0.0 & 0.5\\ 
\hline
10 & 1 &  5.0 & 0.5 & 0.1\\ 
\hline
\end{tabular}
\end{center}

\begin{center}
\begin{tabular}{ | m{1.5cm} || m{1cm} | m{1cm}| m{1cm} | m{1cm}| } 
\hline
\multicolumn{5}{|c|}{104/110 MR-coded BCJR-once rate} \\
\hline
\hline
 Traces & $\beta_b$ &  $\beta_e$ & $\beta_i$ & $\beta_0$\\
  \hline
1 & 1  & 0.5 & 0.0 & 1.0\\
\hline
2 & 1  & 0.1 & 0.0 & 1.0\\ 
\hline
4 & 1 &  0.1 & 0.0 & 0.5\\ 
\hline
6 & 1 &  0.1 & 0.0 & 0.5\\ 
\hline
8 & 1 &  0.5 & 0.5 & 0.5\\ 
\hline
10 & 1 &  5.0 & 0.5 & 1.0\\ 
\hline
\end{tabular}
\end{center}

\begin{center}
\begin{tabular}{ | m{1.5cm} || m{1cm} | m{1cm}| m{1cm} | m{1cm}| } 
\hline
\multicolumn{5}{|c|}{100/110 MR-coded Hamming distance} \\
\hline
\hline
 Traces & $\beta_b$ &  $\beta_e$ & $\beta_i$ & $\beta_0$\\
 \hline
1 & 1  & 0.5 & 0.0 & 0.5\\
\hline
2 & 1  & 0.5 & 0.0 & 0.5\\ 
\hline
4 & 1 &  0.5 & 0.0 & 0.5\\ 
\hline
6 & 1 &  0.5 & 0.1 & 0.5\\ 
\hline
8 & 1 &  0.5 & 0.1 & 0.5\\ 
\hline
10 & 1 &  5.0 & 0.0 & 0.5\\ 
\hline
\end{tabular}
\end{center}

\begin{center}
\begin{tabular}{ | m{1.5cm} || m{1cm} | m{1cm}| m{1cm} | m{1cm}| } 
\hline
\multicolumn{5}{|c|}{100/110 MR-coded BCJR-once rates} \\
\hline
\hline
 Traces & $\beta_b$ &  $\beta_e$ & $\beta_i$ & $\beta_0$\\
 \hline
1 & 1  & 1.0 & 0.0 & 1.0\\
\hline
2 & 1  & 0.1 & 0.0 & 1.0\\ 
\hline
4 & 1 &  0.5 & 0.0 & 0.5\\ 
\hline
6 & 1 &  0.5 & 0.1 & 0.5\\ 
\hline
8 & 1 &  0.5 & 0.1 & 0.5\\ 
\hline
10 & 1 &  0.5 & 0.5 & 1.0\\ 
\hline
\end{tabular}
\end{center}

\end{document}